\documentclass[aps,prl,twocolumn,showpacs,superscriptaddress,groupedaddress]{revtex4}  
\usepackage{graphicx}  
\usepackage{dcolumn}   
\usepackage{bm}        
\usepackage{amssymb}   
\usepackage{amsmath}
\begin{document}
\title{Transverse  Acoustic Phonon Transistor Based on Asymmetric Potential Distribution}

\author{H. \surname{Jeong}}

\affiliation{School of Information and Communications, Gwangju Institute of Science and Technology,
Gwangju 500-712, Korea}

\author{Y. D.  \surname{Jho}}
\email{jho@gist.ac.kr} \affiliation{School of Information and Communications, Gwangju Institute of
Science and Technology, Gwangju 500-712, Korea}

\author{C. J. \surname{Stanton}}
\affiliation{Department of Physics, University of
Florida, Gainesville, Florida 32611, USA}



\date{\today}

\begin{abstract}
We experimentally demonstrate a transverse acoustic (TA) phonon transistor. Phonons are coherently initiated by femtosecond photocarrier screening on potential gradients. Although translational symmetry within the isotropic plane normally prohibits optical generation of TA phonons, we show that the combined application of an external bias in the vertical \textit {and} lateral directions can break the selection rules, generating the forbidden TA mode. The amplitude and on-state time of the TA mode can be modulated by the external field strength and size of the laterally biased region. The observed frequency shift with an external bias as well as the strong geometrical dependence confirm the role of the asymmetric potential distribution in electrically manipulating the crystal symmetry to control and activate the transistor.
\end{abstract}


\pacs{} \maketitle

The ability to tailor materials and device elements below the phonon mean free path in such a way that the vibrational/phononic properties can be modified and precisely controlled has gained widespread interest in recent years \cite{Maldovan}. Some applications include controllability of the thermal conductivity for on-chip heat management \cite{Yu}, charge transfer via acoustic (AC) pulses \cite{Young}, and the realization of phonon lasers \cite{Fainstein}. The acoustic counterpart of diodes \cite{Li} and metamaterials \cite{Guenneau} open further opportunities for the prospective development of phononics.

The coherent vibrational properties can typically be investigated using transient coupling between electrons and dynamic strains under femtosecond laser excitation. Such techniques have been reported for multiferroic crystals \cite{Ruello}, zinc-blende crystals \cite{Matsuda}, wurtzite semiconductors \cite{Lin,Chern}, and their mixtures \cite{Matsuda2008}. Among a vast range of crystalline structures, wurtzite heterostructures revealed strongly enhanced electron-AC phonon coupling efficiency for both zone-folded and propagating AC phonons owing to the large internal polarization fields at pseudomorphically strained interfaces \cite{Chern}. The small absorption depth ($\leq$100 nm) therein compared to the conventional excitation spot size ($\geq$10 $\mu$m) provides ideal plane waves propagating along the growth direction; in the anisotropic plane, transverse acoustic (TA) and longitudinal acoustic (LA) phonons are simultaneously observed \cite{Chen}, whereas in isotropic systems grown along the polar $c$-axis, only the LA mode is allowed \cite{Lin}.

The role of electric fields alongside the growth axis in generating optical phonons \cite{Dekorsy} and AC phonons \cite{Lin,Chen} has long been identified. By positing the growth direction as the decisive factor fixing the phonon modes, however, previous efforts neglected external manipulation of the crystalline symmetry. In analogy with an electrical transistor, implemented under combined potential gradients along the vertical and lateral directions, our approach in this work is to simultaneously manipulate the normal and shear strain distributions in a mutually correlated fashion to attain an unprecedented degree of freedom in amplifying and switching the forbidden TA modes in an isotropic wurtzite system.

\begin{figure}[!t]
\centering
\includegraphics[scale=0.14,trim=20 33 30 50]{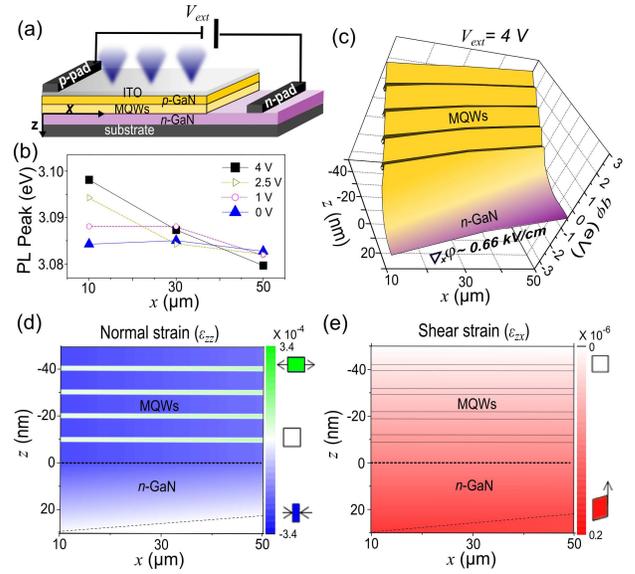}
\caption{ (a) Schematic of structure with asymmetry potential distribution; blue cones denote laser excitation spots moving along the $x$-axis (from $p$-contact to $n$-contact). (b) PL peak energy as a function of excitation position in the lateral ($x$) direction under external bias ($V_{ext}$). (c) Conduction band profile near MQWs/$n$-GaN interface calculated from the PL peak energy variations at 4 V. Spatial distributions of (d) normal and (e) shear strains deduced from the conduction band profile.}
\end{figure}
A representative sample used here under asymmetric potential distribution at room temperature is illustrated in Fig. 1(a). GaN-based multiple quantum wells (MQWs) acting as LA epicenters \cite{Lin} were sandwiched between $p$- and $n$-regions along the $c$-axis ($\equiv$ $z$-axis) so that the external bias $V_{ext}$ induces vertically exerted fields compensating for the piezoelectric field $E^P_z$ \cite{Jho2002}. The fundamental importance of the structure lies in the additional formation of \textit{lateral} electric fields by interdigitated contact pads; the sample has narrow $p$- and $n$- electrodes, which are laterally spaced by $\sim$80 $\mu$m in the $x$ direction on top of an abnormally-thin indium tin oxide (ITO) layer ($\sim$40 nm) to spread a nonuniform electric current density along the $x$-axis. The following layers were sequentially grown on a sapphire substrate by metal–organic chemical vapor deposition: 3-$\mu$m-thick undoped GaN, 2.5-$\mu$m-thick $n$-GaN, six quantum well (QW) layers of 2-nm-thick In$_{0.1}$Ga$_{0.9}$N encased by seven 8-nm-thick GaN barriers, 120-nm-thick $p$-Al$_{0.05}$Ga$_{0.95}$N, and 250-nm-thick $p$-GaN. The doped electron $n_e$ and hole concentrations $n_p$ were estimated to be 7 $\times$10$^{17}$ cm$^{-3}$ and 8$\times$10$^{17}$ cm$^{-3}$, respectively. To determine the potential distributions in both the vertical and lateral directions, the spatially-resolved photoluminescence (PL) was measured, as shown in Fig. 1(b), under excitation by a frequency-doubled Ti:sapphire laser at 367.5 nm onto a spot size of $\sim$10 $\mu$m.

As the excitation spot was moved away from the $p$-electrode in the lateral $x$ direction in Fig. 1(b), the PL peak energy was red-shifted when $V_{ext}$ was applied. The red shift of the PL peak energy, which was distinctive from those reported for vertically strained QWs, indicates that the actual magnitude of the applied bias along the $z$-axis differed from $V_{ext}$ and decreased along the $x$-axis in terms of the quantum-confined Stark effect \cite{Jho2002}. From variational calculations of the PL peak energy, we estimated $E^P_z$ to be 0.98 MV/cm and determined the laterally varying potential profile $\varphi(x,z)$. The resultant conduction band profiles are evaluated in Fig. 1(c) at a $V_{ext}$ of 4 V across the MQW and $n$-depletion regions. By following the procedures introduced above, the lateral electric field $E_{x}=-\nabla_{x}\varphi$ was found to increase with $V_{ext}$, reaching $\sim$0.66 kV/cm at 4 V at the end of the depletion region. In Fig. 1(c), $E^P_z$ was almost fully compensated at 4 V by the vertically applied fields; the magnitude of the net electric field along the $z$-axis, $E_{z}=-\nabla_{z}\varphi$, was on the same order as $E_{x}$ within the QWs.

\begin{figure}[!t]
\centering
\includegraphics[scale=0.1,trim=35 30 0 0]{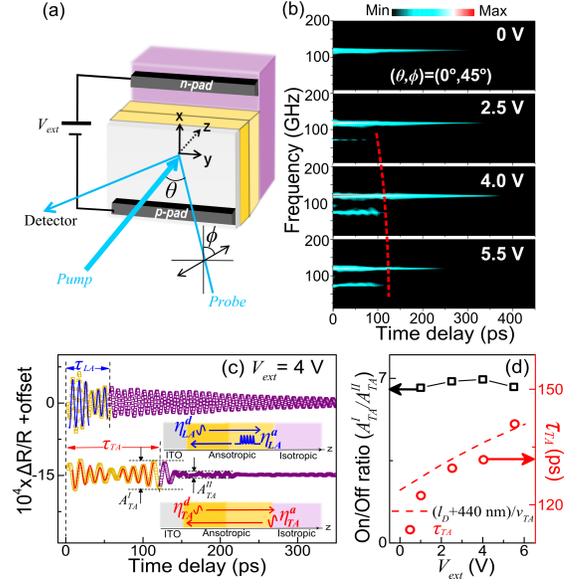}
\caption{ (a) Measurement schematic for differential reflectivity spectra due to AC phonons. (b) Contour of spectrally decomposed amplitudes of dynamic Fabry-Perot interference under different $V_{ext}$ from 0 to 5.5 V. (c) Band-pass-filtered signals in LA (top) and TA (bottom) modes. Insets outline the epicenters and propagation of the AC modes. (d) On/off amplitude ratio of TA wave and on-state time $\tau_{TA}$ as a function of $V_{ext}$.}
\end{figure}

On the basis of the piezoelectric tensor $d_{nm}$, the strain components in hexagonal symmetry in wurtzite GaN are coupled with the electric field distributions as $\varepsilon_{m}$ = $\sum\limits_{n=1}^3 d_{nm}E_{n}$ ($m$ = 1,2,...,6) \cite{Nye}, where nonzero values of $E_{3}=E_{z}$ and $E_{1}=E_{x}$ lead to the normal ($\varepsilon_{zz}=\varepsilon_{3}$) and shear strain components ($\varepsilon_{zx}=\varepsilon_{5}$) in the $x-z$ plane, respectively. The normal strain $\varepsilon_{zz}$ calculated in Fig. 1(d) was compressive in the barriers and weakly tensile in the QWs at 4 V. On the other hand, the shear component ($\varepsilon_{zx}$) in Fig. 1(e) became prominent throughout the biased region and reached a maximum value near the end of the $n$-depletion region. When coherent AC phonons are displacively initiated along the $z$-axis via selective screening by photocarriers on the strained lattice, the modal driving force in the loaded string model \cite{Chern,Stanton2005} could be correlated with the strain as
\begin{equation}
S_i(x,z,t)=\frac{1}{\rho_{0}}\bar{C}_i\varepsilon_i(x,z)H(t),
\end{equation}
where the mode index $i$ corresponds to either LA or TA, $\bar{C}_{i}$ is the effective elastic constants \cite{Supplementary}, $\varepsilon_{LA}$ ($\varepsilon_{TA}$) corresponds to the normal (shear) strain $\varepsilon_{zz}$ ($\varepsilon_{zx}$), and $H$$(t)$ is the Heaviside step function describing the instantaneous excitation of the driving forces by transient field screenings.

To investigate the interplay between the asymmetric potential distribution and the modal AC phonon dynamics, we measured the time-resolved differential reflectivity spectra (DRS) as sketched in Fig. 2(a). The fluence of the pump beam was $\sim$85 $\mu$J/cm$^{2}$ in order to abruptly screen out the potential gradients. The incident angle $\theta$ of the pump beam for phonon generation was fixed at zero, whereas $\theta$ for the polarized probe beam for phonon detection was set to less than 5$^{\circ}$ unless otherwise mentioned (Fig. 5). For photocarrier excitation in the MQW and depletion regions, the pump and probe beam energies were degenerate at 367.5 nm, with a penetration depth $\xi$ of around 700 nm. The probe polarization $\phi$ was fixed at 45$^{\circ}$ with respect to the incident plane except in Fig. 4, whereas the pump beam was polarized perpendicular to the probe beam.

The transient oscillatory components in DRS, $\Delta R/R$, were analyzed to produce contour plots of the modal amplitudes in Fig. 2(b). Each frequency mode of the oscillation was caused by dynamic Fabry-Perot interference between the probe beams reflected from the surface and from the wavefront of the propagating strain waves. The modal frequency is expressed as $f_{i}$=$2v_{i}n$cos($\theta_{t}$)/$\lambda$$_{probe}$, where $\theta_{t}$ is the angle of the probe transmission inside the material, $n$ is the refractive index, and $\lambda$$_{probe}$ is the wavelength of the probe beam. Not only the LA frequency $f_{LA}$ $\sim$120 GHz, but also new spectral component at $f_{TA} \sim$70 GHz, emerged with $V_{ext}$. Further, $f_{LA}$ ($f_{TA}$) matches well the velocity $v_{LA}$ ($v_{TA}$) of $\sim$7300 m/s (4200 m/s) \cite{Supplementary}. The actively induced TA mode under a nonzero $V_{ext}$ could not be explained by any previous investigations of the isotropic plane. Furthermore, the TA waves abruptly disappeared around $\tau_{TA}$ ($\sim$130 ps at 4 V), which is indicated by a curved dashed line in Fig. 2(b).

The detailed lineshape analysis in Fig. 2(c) further revealed that the TA and LA signals (scattered lines) could be matched with the solid fitting curves by adaptively integrating the product of the sensitivity function $\digamma_i$ and the strain $\eta_i$; i.e., $\Delta R/R=\int_{-\infty}^{\infty} dz \digamma_i e^{-z/\xi} \cdot \eta_i $, where $\eta_i$ is the linear supposition between descending waves ($\eta_{i}^{d}$, which propagate toward the substrate and represent decaying signals for $0<t<\tau_{i}/2$), and ascending waves ($\eta_{i}^{a}$, which propagate toward the surface and represent growing signals for $\tau_{i}/2<t<\tau_{i}$). By taking the time scale in the biased region ($\tau_{LA}$ $\sim$ 55 ps and $\tau_{TA}$ $\sim$ 130 ps at 4 V) into account, both $\eta_{TA}^{d}$ and $\eta_{LA}^{d}$ were determined to spatially originate from the middle of the surface depletion region (SDR), whereas $\eta_{TA}^{a}$ and $\eta_{LA}^{a}$ were found to have different epicenters, in the $n$-depletion and MQW regions, respectively. These spatially concentrated AC strains, therefore, could be simplified into $\eta_{i}^{a}$=$\int_{D_i^{a}}\varepsilon_{i}(z)dz\cdot\delta(z+v_{i}t)$ for the ascending waves and $\eta_{i}^{d}$=$\int_{D_i^{d}}\varepsilon_{i}(z)dz\cdot\delta(z-v_{i}t)$ for the descending waves, where the region of integration $D_i^{d}$  ($D_i^{a}$) corresponds to the SDR (either to the MQW region for the LA mode or to the $n$-depletion region for the TA mode).
\begin{figure}[!t]
\centering
\includegraphics[scale=0.12,trim=35 35 0 50]{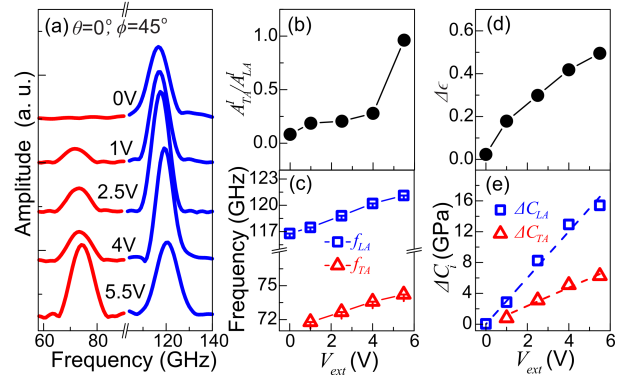}
\caption{ (a) Modal spectra for ascending waves as a function of $V_{ext}$. (b) TA amplitude normalized by LA mode with increasing $V_{ext}$. (c) Peak frequency for each AC mode under $V_{ext}$. (d) $V_{ext}$-induced variation in dielectric constant. (e) $V_{ext}$-induced elastic constant variation for each mode.}
\end{figure}

The on/off ratio of the TA mode, $A^I_{TA}/A^{II}_{TA}$, was held at $\sim$7 with an external bias in Fig. 2(d), where $A^I_{TA}$ ($A^{II}_{TA}$) corresponds to the maximum TA amplitude for the on-state (off-state) for time domain $I$ ($II$) with $t<\tau_{i}$ ($\tau_{i}<t<2\tau_{i}$). This is explained by the estimated AC reflectivity of about 0.19 due to the AC impedance mismatch at the ITO/$p$-GaN interface. The increasing on-state time $\tau_{TA}$ (red circles) in Fig. 2(d), on the other hand, was in good agreement with the calculated travel time (dashed line) between the end of the $n$-depletion region and the SDR. In this regard, the slightly increased value of $\tau_{TA}$ between 2.5 and 5.5 V ($\sim$11 ps) can be converted into the elongated $n$-depletion region width $l_D(n_e,n_p,V_{ext})$ \cite{Jho2002}. It was previously postulated that the dielectric tensor modulation due to transient shear strain waves can be detected only with oblique probe incidence in an optically isotropic medium \cite{Matsudaoff}. Therefore, the digitized appearance of the TA mode even under normal incidence in Fig. 2, which matches exactly the time-of-flight in the laterally biased region, implies that the hexagonal symmetry was broken there.

To precisely trace the frequencies and amplitudes without phase-change-induced errors at the AC interfaces \cite{Supplementary}, the Fourier amplitudes for $\eta_{i}^{a}$ integrated over $\tau_{i}/2<t<\tau_{i}$ are presented for the TA (red lines) and LA modes (blue lines) at different $V_{ext}$ in Fig. 3(a). To illustrate the amplification of the TA mode, the TA signal was normalized by that of the persistently manifested LA mode as $A^I_{TA}/A^I_{LA}$ in Fig. 3(b). The TA amplitude increased consistently with increasing shear strain even beyond 4 V, at which electronic tunneling overrides the strain increment and causes a relative decrease in the LA signals. Photocarrier screening within QWs (referred to as in-well screening, on the order of the exciton Bohr radius $a_B$) generally plays a major role in LA mode generation \cite{Chern}. However, this fast electronic transport could not weaken the TA amplification in Fig. 3(a, b), which suggests a distinctive origin of TA mode generation at a much larger scale than $a_B$. Another intriguing feature regarding the electrically broken symmetry was the spectral blue shift in both modes under increasing $V_{ext}$ in Fig. 3(a). The peak frequencies of the LA and TA modes increased with the bias by $\sim$1.7\%, as traced in Fig. 3(c). In terms of the birefringence in the anisotropic region, we firstly examined the $V_{ext}$-dependent refractive index change, $n=\sqrt{\epsilon^{o}+\Delta \epsilon (V_{ext})}$, by using ellipsometry in Fig. 3(d). The unperturbed dielectric constant $\epsilon^{o}$ was evaluated to be 8.64 at 367.5 nm at zero bias. Then, inserting the measured $n$ into $f_i$ under $V_{ext}$, we extracted $ v_{i}$, given by $v_i=\sqrt{ \frac{C_i+\Delta C_i (V_{ext})}{\rho_0}} $. The additional bias-dependent changes, $\Delta \epsilon (V_{ext})$ and $\Delta C_i (V_{ext})$, are presented in Fig. 3(d, e) and could be further compared with the analytical expressions through the birefringence and strain-induced nonlinear elasticity in the anisotropic region \cite{Supplementary}.

\begin{figure}[!t] \centering
\includegraphics[scale=0.18,trim=35 25 0 80]{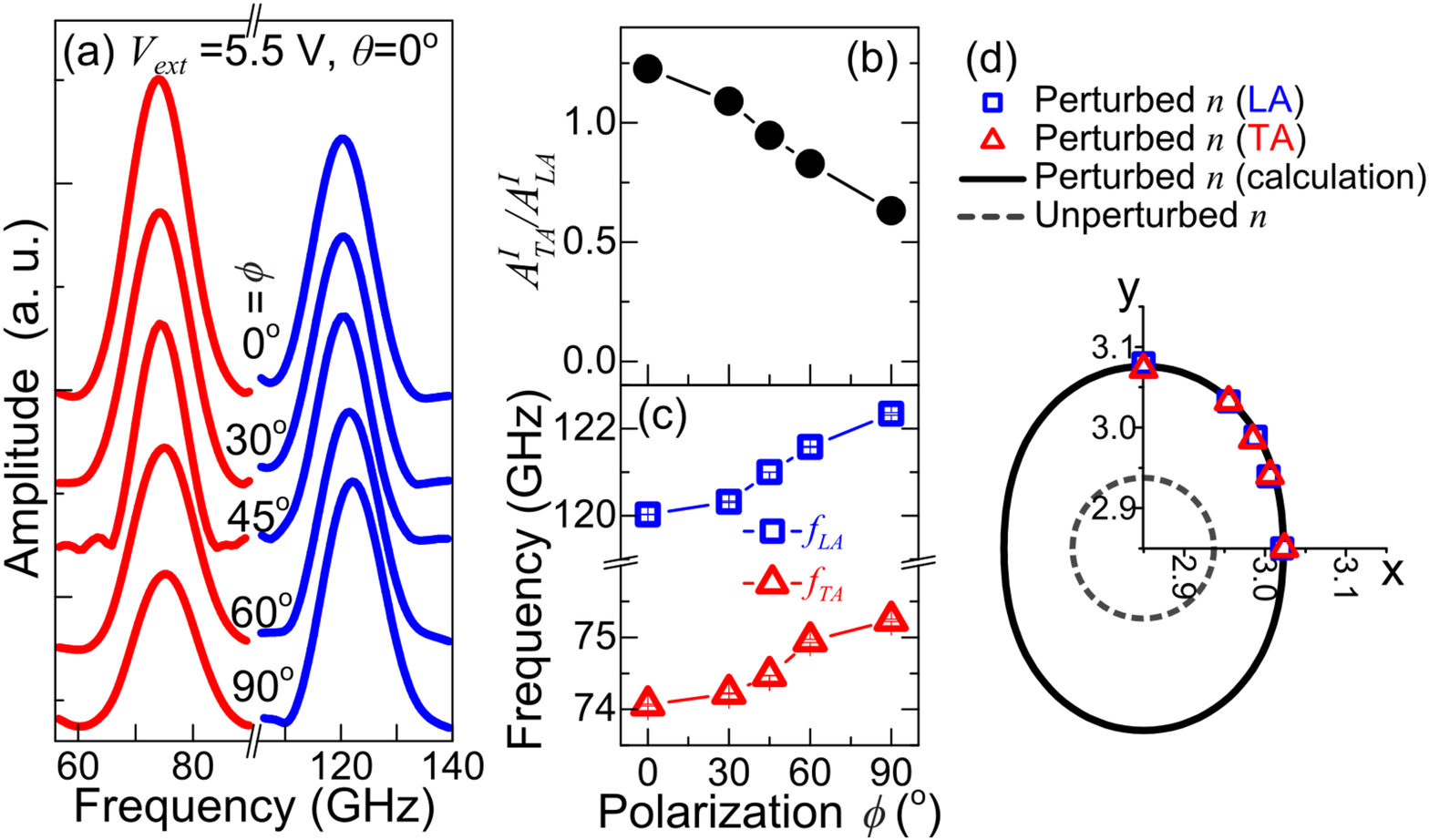} \caption{
(a) Modal spectra for ascending waves at different probe polarizations $\phi$ under $V_{ext}$ of 5.5 V. (b) TA amplitude normalized by LA mode with $\phi$ from $0^{\circ}$ to $90^{\circ}$. (c) Peak frequency for each AC mode as a function of $\phi$. (d) Refractive index ellipsoids for perturbed (at 5.5 V) and unperturbed structure (at 0 V).}
\end{figure}
\begin{figure}[!t]
\centering
\includegraphics[scale=0.091,trim=35 120 0 100]{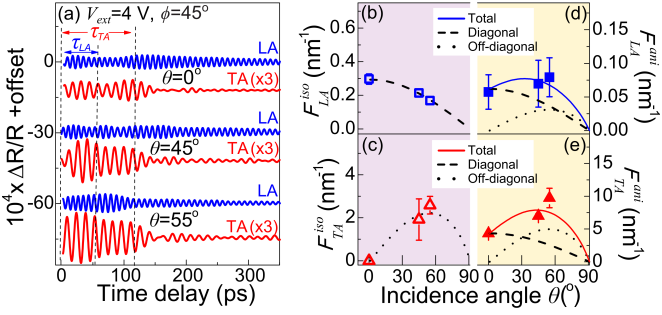}
\caption{ (a) Time domain signals for different probe incidence angles. Optical sensitivities of AC modes as a function of incident angle in the isotropic (b, c) and anisotropic (d, e) regions.}
\end{figure}

Regarding the electrically manipulated crystal symmetry, $E_x$ possibly distorted the hexagonal symmetry via the third-order elastic (TOE) tensors into monoclinic symmetry \cite{Fuck}. Concretely, the tilting of the principal axis away from the $c$-axis by $E_x$-induced monoclinic deformation could be illustrated by angular perturbations of the elastic constants and refractive index in the $x-y$ plane. Figure 4(a) shows the modal spectra at a $V_{ext}$ of 5.5 V, integrated over $\tau_{i}/2<t<\tau_{i}$, as we rotate the probe polarization $\phi$ from the $x$-axis. Two noteworthy observations were made: (1) in clear contrast to the constant LA mode (blue lines), the TA amplitude (red lines) monotonically decreased with $\phi$ in Fig. 4(a), indicating that the TA mode was partially polarized along the same direction as $E_x$. Indeed, the TA magnitude normalized by the LA mode ($A^I_{TA}/A^I_{LA}$) in Fig. 4(b) revealed a drastic decrease by about 50\%. (2) Despite the fixed $V_{ext}$, the modal frequencies $f_{LA}$ and $f_{TA}$ exhibited very similar increases with $\phi$, as shown in Fig. 4(c), owing to the birefringence in the anisotropic region. As a function of $\phi$, in this regard, $n$ was extracted either from $f_{LA}$ (blue squares) or from $f_{TA}$ (red triangles) in Fig. 4(d). The dashed inner circle in Fig. 4(d) displays the isotropic refractive index ellipsoid without $E_x$, whereas the solid outer circle exhibits the values calculated via the $E_x$-induced birefringence. In a comparative structure without $E_x$ even under an applied $V_{ext}$, where the TOE tensor has preserved hexagonal symmetry, this $V_{ext}$-induced birefringence and nonlinear elasticity vanished \cite{Supplementary}.

To gain a more quantitative insight into the relationship between $\digamma_i$ and the axial symmetry, we selectively measured the $s$-polarized component of the probe reflectance as a function of $\theta$ at 4 V, as shown in Fig. 5. The probe polarization angle $\phi$ was fixed at 45$^{\circ}$ to maximize the sensitivity of the $x$-polarized TA mode \cite{Matsuda2008}. As a result, more complicated AC propagation dynamics appear in Fig. 5(a); the TA mode signals monotonically increased with $\theta$ not only in the purely anisotropic region ($I$: $t<\tau_i$) but also in the purely isotropic region ($III$: $2\tau_{i}<t<3\tau_{i}$). We note that the abruptly suppressed TA mode at $\tau_{TA}$ was particularly prominent at $\theta$=$0^{\circ}$ in Fig. 5(a). In contrast, the LA amplitude with $\theta$ varied in different time domains, slightly increasing up to $\tau_{LA}$ and then rapidly decreasing after 2$\tau_{LA}$. In the time domain $II$, the increasing and decreasing tendencies were mixed for LA waves.

The $\theta$-dependent $\digamma_i$ was inferred from a model which correlates the spatial dynamics in different symmetry regions with the time domain signals. The modal amplitudes $A_{i}^{j}$ at $f_i$ in the time domain $j$ were decomposed into ascending and descending parts, from which the relative values of $\eta_i^a$, $\eta_i^d$, and the sensitivities in the anisotropic region, $\digamma_{i}^{ani}$ for $z \leq v_i \cdot \tau_i$, and isotropic region, $\digamma_{i}^{iso}$ for $z \geq v_i \cdot \tau_i$, could be extracted in Fig. 5(b-e). The analytic expressions for $\digamma_{i}$ have been discussed separately for isotropic \cite{Matsudaoff} and anisotropic crystals \cite{Pezeril}, according to the perturbations of the dielectric tensor due to strain. In isotropic materials, only the diagonal (off-diagonal) components in the perturbed dielectric tensor are induced by the LA (TA) waves. That is why the opposite tendencies are predicted for $\digamma_{LA}^{iso}$ (dashed line) and $\digamma_{TA}^{iso}$ (dotted line) within our range of $\theta$ values, in agreement with the experimental values (scattered lines) in Fig. 5(b, c). To calculate $\digamma_{i}^{ani}$, it is necessary to consider the mixed nature of the dielectric tensor modulations resulting from both diagonal and off-diagonal perturbations. In Fig. 5(d, e), the experimentally obtained values of $\digamma_{i}^{ani}$ were reproduced by linear combinations of the diagonal perturbations in the dielectric tensor ($\propto \digamma_{LA}^{iso}$) and off-diagonal perturbations in the dielectric tensor ($\propto \digamma_{TA}^{iso}$). Most importantly, the digitized appearance of the typically forbidden TA mode at $\theta=0^{\circ}$ (Fig. 2) was theoretically verified by comparing the zero $\digamma_{TA}^{iso}$ [Fig. 5(c)] and nonzero value of $\digamma_{TA}^{ani}$ [Fig. 5(e)].

In summary, we reported a prototypical AC transistor by which one can amplify and switch on and off TA waves via a combination of lateral and vertical potential gradients. By applying lateral electric fields in the isotropic plane, we could modify the selection rules, allowing the generation of the forbidden TA mode. Owing to the anisotropic nature of the elastic tensor over the laterally biased scale, the frequency, on-state times, and sensitivities of the TA waves were also externally or axially modulated. Finally, we note that the asymmetric potential distribution described in this Letter could also contribute to the active control of optical phonon modes, consistent with crystal symmetry. Furthermore, as the electric fields couple strongly to the phonons in piezoelectric nanostructures, as demonstrated in this Letter, the activation of phononic functionalities with a degree of control analogous to that for manipulating electrons in transistors could be heralded as the essential step in advancing integrated phononic circuits for logical processing.\\
We acknowledge useful discussions with P. Ruello, J. Kono, P. Gr\"{u}nberg, and K. J. Yee. This work was funded by the Basic Science Research Program through the National Research Foundation of Korea (NRF-2012-042232; 2013-068982). The samples used in this work were provided from LG electronics and Korean Photonics Technology Institute.


\begin{thebibliography}{sorting}



\bibitem{Maldovan} See, e.g., M. Maldovan, Nat. \textbf{503}, 209 (2013).

\bibitem{Yu} J.-K. Yu, S. Mitrovic, D. Tham, J. Varghese, and J. R. Heath, Nat. Nanotechnol. \textbf{5}, 718 (2010).


\bibitem{Young}E. S. K. Young, A. V. Akimov, M. Henini, L. Eaves, and A. J. Kent, Phys. Rev. Lett. {\bf 108}, 226601 (2012).





\bibitem{Fainstein} See, e.g.,  A. Fainstein, N. D. Lanzillotti-Kimura, B. Jusserand, and B. Perrin, Phys. Rev. Lett. {\bf 110}, 037403 (2013).



\bibitem{Li} X.-F. Li, X. Ni, L. Feng, M. H. Lu, C. He, and Y. F. Chen, Phys. Rev. Lett. {\bf 106}, 084301 (2011).


\bibitem{Guenneau} S. Guenneau, A. Movchan, G. P\'{e}tursson, and S A. Ramakrishna,  New. J. Phys. {\bf 9}, 399 (2007).

\bibitem{Ruello} M. Lejman, G. Vaudel, I. C. Infante, V. E. Gusev, B. Dkhil, and P. Ruello, Nat. Commun. {\bf 5}, 5301 (2014).





\bibitem{Matsuda} O. Matsuda, O. B. Wright, D. H. Hurley, V. E. Gusev, and K. Shimizu, Phys. Rev. Lett. {\bf 93}, 095501 (2004).











\bibitem{Lin} K. H. Lin, C. M. Lai, C. C. Pan, J. I. Chyi, J.W. Shi, S. Z. Sun, C. F. Chang, and C.-K. Sun, Nature Nanotech. {\bf 2}, 704 (2007).













\bibitem{Chern} G-.W. Chern, C.-K. Sun, G. D. Sanders, and C. J. Stanton, in
\textit{Ultrafast Dynamical Processes in
Semiconductors}, Topics Appl. Phys. {\bf 92}, 339
(2004).

\bibitem{Matsuda2008} O. Matsuda, O. B. Wright, D. H. Hurley, V. E. Gusev, and K. Shimizu, Phys. Rev. B {\bf 77}, 224110 (2008).


\bibitem{Chen} C.-C. Chen, H.-M. Huang, T.-C. Lu, H.-C. Kuo, and C.-K. Sun, Appl. Phys. Lett. {\bf 100}, 201905 (2012).



\bibitem{Dekorsy}  T. Dekorsy, G. C. Cho and H. Kurz, in \textit{Light
Scattering in Solids VIII}, edited by M. Cardona and G. G\"{u}ntherodt {\bf 76}, pp. 169-209 (Springer, Berlin, 2000).















\bibitem{Jho2002} Y. D. Jho, J. S. Yahng, E. Oh, and D. S. Kim, Phys. Rev. B {\bf 66}, 035334 (2002).

\bibitem{Nye} J. F. Nye, \textit{Physical Properties of Crystals} (Oxford University Press, Oxford, 1985).

\bibitem{Stanton2005} R. Liu, G. D. Sanders, C. J. Stanton, C. S. Kim, J. S. Yhang, Y. D. Jho, K. J. Yee, E. Oh, and D. S. Kim, Phys. Rev. B {\bf 72}, 1 (2005).


\bibitem{Supplementary}
See Supplementary Material for the excitation fluence
dependence of phonon amplitudes, wavelength-dependent
modulations of differential reflectivity spectra,
detailed envelope lineshapes of different modes, lateral-electric-field-induced mechanical and optical anisotropy, and
the method for extracting sensitivities.













\bibitem{Fuck} R. F. Fuck and I. Tsvankin, Geophysics {\bf 74}, WB79 (2009).

\bibitem{Matsudaoff} O. Matsuda and O. B. Wright, Anal. Sci. {\bf 17}, S216 (2001).

\bibitem{Pezeril} T. Pezeril, P. Ruello, S. Gougeon, N. Chigarev, D. Mounier, J.-M. Breteau, P. Picart, and V. Gusev, Phys. Rev. B {\bf 75}, 174307 (2007).





\end{thebibliography}
\end{document}